\documentclass[preprint]{aastex}

\begin{document}

\title{\emph{HST}/NICMOS Observations of the Embedded Cluster in \\NGC 2024:
Constraints on the IMF and Binary Fraction}
\author{Wilson M. Liu, Michael R. Meyer, Angela S. Cotera, and Erick T. Young}
\affil{Steward Observatory, University of Arizona}
\affil{933 N. Cherry Ave., Tucson, AZ, USA 85721}
\email{wliu@as.arizona.edu, mmeyer@as.arizona.edu, acotera@as.arizona.edu,
 eyoung@as.arizona.edu}

\begin{abstract}
We present an analysis of NICMOS observations of the embedded cluster
associated with NGC 2024.  An analysis of the cluster color-magnitude
diagram (CMD) using the models of \citet{DM98} and
\citet{BCAH98} indicates that the ratio of intermediate
mass (1.0 to 10.0 $M_{\sun}$) to low mass (0.1 to 1.0 $M_{\sun}$) stars is
consistent with the stellar initial mass function (IMF) for the field.
In addition to the CMD analysis, we present results on the multiplicity of
stars in the region.  Three companions (in a sample of 95 potential
primaries) were found, with angular separations between $0\arcsec .4$ and
$1\arcsec .0$, translating to a projected linear separation of 184 AU to
460 AU for an estimated distance of 460 pc. The completeness of
binary detections is assessed using recovery fractions calculated
by a series of tests using artificially generated companions to
potential primaries in the data frames.  We find that the
binary fraction in NGC 2024 is consistent with that of \citet{DM91}
for solar neighborhood stars over the range of separations and
companion masses appropriate for our survey.
\end{abstract}

\section{Introduction}

Two fundamental questions regarding star formation concern the
universality of the stellar initial mass function (IMF) and the fraction of stars
which exist in multiple systems at birth.  More specifically, it is not known
whether the functional form of the IMF is dependent on the local characteristics
of a star-forming environment, or if it is universal, unchanging between different
regions.  A similar question exists regarding the fraction of stars observed to be in
multiple systems in a star-forming region: is there a correlation between this
binary fraction and the characteristics of star-forming regions, such
as cluster age and central stellar density?  If so, what is this correlation?
NGC 2024, a young embedded cluster associated with an H II region in the Orion B
giant molecular cloud provides a good opportunity to investigate both of these
aspects of star formation.  The relatively high density of the cluster compared
to other star-forming regions allows us to study a statistically significant
sample in a small field-of-view, while the extinction through the associated cloud core
is high enough to eliminate contamination from background sources.  Additionally, the
high resolution of the data taken with the Near-Infrared Camera and Multi-Object
Spectrometer (NICMOS) onboard the \emph{Hubble Space Telescope} (\emph{HST})
presented in this study allows us to study the cluster at higher spatial resolution
in the J and H-bands than previously done.  These data are also complimentary
to previous observations of the embedded cluster in NGC 2024 done in the I-band with
the Wide Field Planetary Camera 2 onboard HST \citep{PSG97} and in the K-band with
adaptive optics \citep[hereafter BSC03]{bsc}.

The IMF of several nearby star-forming regions has been well studied in
recent years.  The first study which investigated the field star IMF over a large
range of masses ($0.1 M_{\sun}$ to $~10 M_{\sun}$) was \citet{MS79} and has been
subsequently updated by \citet{Scalo86} and
\citet{K01}, among others.  Previous investigation into the IMF in young
clusters in Orion include the near-IR study of the Trapezium by \citet{Luhman} as well
as studies of NGC 2024 including those by \citet{CRR96} and \citet{Meyer_phd}.
These studies, along with studies of other star-forming regions, have shown
evidence that the IMF is universal, unchanging regardless of local environmental
conditions \citep{Meyer_ppiv}.  Our study of NGC 2024 will assess the similarity of
the cluster IMF to those derived in previous studies
by comparing the ratio of intermediate ($1.0 - 10 M_{\sun}$) to low
($0.1 - 1.0 M_{\sun}$) mass objects in NGC 2024 to the ratio expected from
IMFs derived in \citet{K01} for the field.  This ratio can be determined by using
near-infrared photometry of the cluster which is then compared to theoretical
models of pre-main sequence (PMS) objects, provided an estimate for the
cluster age is available.  As a diagnostic for the IMF,
the primary advantages of using this ratio lies with the fact that the mass
bins are wider than the errors in stellar mass determination from PMS models, hence
the value of the ratio should be robust even with these uncertainties.  However,
the use of wide mass bins also makes this diagnostic insensitive to small
variations in the IMF, allowing one to probe only dramatic differences between
mass distributions.  Nonetheless, by comparing this ratio determined for NGC 2024
to that of the Galactic field, we can probe the possibility of gross differences
in the mass distribution.

The binary frequency of stellar populations has also been the focus of many
recent studies, and an area of particular interest lies in determining
how the binary fraction of a particular region is affected by the local environment.
A study by \citet{DM91} investigated solar neigborhood
stars over a large mass range ($q = M_{comp}/M_{pri} = 0.1$ to 1.0) and
period distribution (1 to $10^{10}$ days).  Other
studies have investigated the binary fraction in star-forming regions.
\citet{GNM93} found the  binary frequency in T Tauri stars in Taurus-Auriga
and Ophiuchus-Scorpius to be $60 \pm 17$ \%, about four times greater
than the binary frequency derived by \citet{DM91} for field stars.
\citet{PSG97} presented the binary fraction for optical sources in
three clusters in Orion: NGC 2024, NGC 2068 and NGC 2071. Using WFPC2 HST
observations, they found the binary fraction to be roughly
equal to that of the field over the separation range studied (138 to 1050 AU).
Although it is not known exactly how the local environment
affects the detected binary fraction in these clusters,
there is evidence that it varies between clusters with different central stellar
densities and ages \citep{KB01}.  By measuring the binary frequency in NGC 2024, we
can gain further insight into the binary frequency of a region which is intermediate
in central stellar density to the sparse T association of Taurus-Auriga and the
rich environment of Trapezium.  It also provides an independent check of the binary
fraction derived for NGC 2024 by other studies, including \citet{PSG97} and
BSC03.

In this study, we present high resolution imaging of NGC 2024 obtained using
NICMOS onboard \emph{HST}.  The high
resolution of the dataset allows us to obtain more accurate photometry of
crowded objects and those embedded in nebulosity.  It also allows better
sensitivity to companions at close separations.  Section \ref{sec-obs}
describes the observations and data reduction procedure and presents
some noteworthy features in the image.  An analysis and discussion of the
color-magnitude diagram (CMD) of the cluster is presented in \S \ref{sec-cmd},
followed by the results for our binary fraction study in \S \ref{sec-bf}.
We summarize and conclude in \S \ref{sec-conc}.

\section{Observations and Data Reduction} \label{sec-obs}

Images of the young embedded cluster associated with NGC 2024 were obtained
on 1998 January 17 using Camera 3 of the NICMOS (NIC3) on \emph{HST}.
Nine images, each $52\arcsec$ by $52\arcsec$, were taken with each filter (F110W and F160W)
in a 3 by 3 mosaic pattern, resulting in a total field of view of $124\arcsec$ by
$124\arcsec$ centered upon 05h 41m 42.4s, -1$\degr$ $55\arcmin$ 15$\arcsec$ .9.  Adjacent
frames overlapped by approximately $8\arcsec$ (15\%).  Integration times
were 96 s for F110W and 80 s for F160W.  These observations were taken as part of
GTO/NIC program 7217 to support grism observations of the same field.

Data reduction was carried out using a combination of IRAF and custom IDL programs.
The images were dark subtracted using artificial dark frames created with
the \emph{nicskydark} routine in the \emph{nicred} package for IRAF \citep{McLeod}.
Each frame suffered from an offset in background level between the quadrants of the
array.  The offsets were determinined in a manner identical to that of
\citet{Luhman} by minimizing the median of the differences of the pixels on the
borders.  Using the top right quadrant as the reference qudrant, the offsets
were applied to the two adjacent quadrants.  The final quadrant was offset to
minimize the difference with the two adjacent frames.  Flat fielding was performed
after correcting the quadrant offsets with flats from the Space Telescope Science
Institute (STScI), also using the \emph{nicred} data reduction package.  Bad
pixels in the frames (i.e., cosmic rays and other artifacts) were located initially
with STScI bad pixel masks, followed by
visual inspection of each frame.  Aberrant pixels were replaced using the
IRAF task \emph{fixpix}, which replaces pixels using a linear interpolation of
surrounding pixels.  The nine images in each band were mosaicked, adjusting the
background offsets for each frame in a manner similar to the quadrant offset
adjustment.  Figure \ref{fig:ngc2024} shows the final mosaicked
two-color image of the central cluster.  Some interesting features and
characteristics of the image are discussed in Section \ref{sec-disc_images}.

\subsection{Photometry}

Ninety-five sources were detected in both the F110W and F160W filters.
Photometry was obtained for 79 of these sources in both bands, the remaining
sources being either too bright (i.e. saturated) or too faint for accurate
photometry.  These sources, as well as the photometry, are presented in Table
\ref{tab:objects}, and their positions are shown in Figure \ref{fig:star_pos}.
Initial detections were made with the IRAF task \emph{daofind} with a $10 \sigma$
detection threshold.  Even with the high initial threshold, each frame produced
many false detections, due to a noisy background and the presence of nebulosity
in the region.  These spurious detections were removed through visual inspection
of each frame.  All final sources were detected in both the F110W and F160W
frames.  Photometry was extracted for each source using the \emph{apphot} routine
in IRAF.  Since NIC3 undersamples the point spread function (PSF)
of the sources, aperture photometry was used for all sources.
Aperture corrections were calculated
using several bright, isolated stars and applied to the photometry to correct
for total signal.  An analysis of aperture corrections versus sky noise
showed the optimal aperture to be 5 pixels for most sources, with the sky
background measured in an annulus from 5-7 pixels around each source.  The
sky annulus was measured as close as possible to the aperture so that the effect
of variablility from nebulosity was minimized.
Apertures and sky annuli were adjusted to obtain accurate photometry of close
companions (see Section \ref{sec-binfrac} for further details about the
binary detections).

Relative astrometry for each source was determined using the
World Coordinate System (WCS) information in the NIC3 header with the \emph{xy2sky}
routine in the WCSTools package \citep{mink}.  The derived coordinates had a fixed
offset with respect to previously determined coordinates (e.g., \citet{Meyer_phd},
among others).  Absolute astrometry was derived by correcting for the offset using the
coordinates in \citet{Meyer_phd} for the brightest sources in the field, IRS1 and
IRS2, as a reference.  The coordinates are presented in Table
\ref{tab:objects} and are accurate to $\lesssim 2\arcsec$, though none of
the analyses in this study are dependent upon absolute astrometry.  Also listed in
Table \ref{tab:objects} are the ID numbers from BSC03, if the object was
detected in both studies.

\subsection{Magnitude Calibration and Color Transformations} \label{sec-magcal}

IRAF magnitudes were calibrated to the Vega system assuming 2.873 $\times 10^{-6}$
and 2.776 $\times 10^{-6}$ Jy/ADU/s and zero magnitude fluxes of 1775 and 1083 Jy for
F110W and F160W respectively.  Detection limits were determined using a series
of artificial star tests.  Artificial stars from 17th to 22nd magnitude in steps of
0.5 mag were added to each of the data frames in both bands using a PSF derived
from several uncrowded sources in areas relatively free of nebulosity.  The
recovery fraction of our detection technique was then assessed.  The 90 \%
competeness limits were found to be $m_{110} \approx 19.5$ mag and $m_{160}
\approx 18.0$ mag.

The NICMOS Vega magnitudes, $m_{110}$ and $m_{160}$, were transformed to the
ground-based CIT system, J and H.  Comparisons of $m_{110}$ vs. J,
$m_{160}$ vs. H, and $(m_{110} - m_{160})$ vs. (J - H) were made
for 21 objects common to this study and a ground-based study of NGC 2024 by
\citet{Meyer_phd}. A linear regession was performed on these relations to obtain
the following color transformations:
\begin{eqnarray}
H = (0.358 \pm 0.080) + (0.935 \pm 0.006) m_{160} \nonumber \\
(J - H) = (-0.734 \pm 0.060) + (1.05 \pm 0.02)(m_{110} - m_{160}) \nonumber
\end{eqnarray}
The $m_{160}$ vs. H, and $(m_{110} - m_{160})$ vs. (J - H)
relations resulted in the smallest error for the linear fit, and were
therefore used to transform the NICMOS magnitudes to the ground-based CIT
magnitudes, with J for each source calculated by using the transformed
H-band magnitude and the (J - H) color: $J = H + (J - H)$.
A J vs. (J - H) CMD of the cluster is
presented in Figure \ref{fig:cmd}.  An analysis and discussion of the CMD will
follow in the next section.

\subsection{Prominent Features in the Image} \label{sec-disc_images}

Given the high resolution of the data, our images show in great detail
the nebulosity and stellar distribution at the center of the cluster.  There
are several features worth noting.
The brightest source on the west is IRS1, while IRS2 is the bright source
on the east. Overall, the image shows a color gradient from east to west in
both the nebulosity and stars, indicating that the east side is more deeply
embedded than the west side.

The nebulosity in the region also shows interesting structure in
places.  The bright knot of nebulosity located about 40" southeast of
the center of the field-of-view (just SE of source 1-21) has several interesting
characteristics.  First, the center of the knot seems to contain an extended
source, possibly a deeply embedded protostellar object.  It is also interesting to
note that the location of the knot corresponds roughly to a compact dust condensation
(FIR4) identified by \citet[1992]{mez88} at 350 and 1300 $\mu m$.  Previous detection
of this emission have also been made at longer near-IR wavelengths (1.6 to 3.6
$\mu m$) by \citet{moo89} and \citet{moo95} and at 450 and 800 $\mu m$ by
\citet{viss}.  Two other FIR objects (FIR2 \& 3)
identified in the \citet[1992]{mez88} are also within the field-of-view of this
study; the nearby sources detected in this study are noted in Table
\ref{tab:objects}.  \citet{cc96} performed observations of these same FIR sources at
98 and 112 GHz and found them to be consistent with extended dust emission
surrounding embedded protostars as well as an outflow for FIR4.

\section{The Color-Magnitude Diagram and Cluster IMF} \label{sec-cmd}

\subsection{Analysis of the CMD} \label{sec-cmd_descrip}

Analyses of the CMD were performed in the natural NICMOS system as well
as the transformed CIT magnitudes.  We sought to derive a ratio
of intermediate (1.0 to $10 M_{\sun}$) to low (0.1 to $1.0 M_{\sun}$) mass stars
in order to compare it to the expected ratio for the field star IMF of \citet{MS79}
and \citet{K01}.  Two sets of models were used for comparison
to the cluster CMD, including the models of \citet{BCAH98} in both NICMOS and
CIT systems, and \citet[updated in 1998; hereafter DM98]{DM98}
in the CIT system, the latter of which covers objects with masses less than
$0.3 M_{\sun}$.  Above this mass, we adopt the models of \citet{DM94}.  Figure
\ref{fig:cmd} shows the cluster CMD in the CIT system along with the
3 $\times\ 10^{5}$ yr isochrone of DM98.  An age of 3 $\times\ 10^{5}$ yr
was adopted for the cluster, as this was determined to be the median age of
embedded objects through an analysis of the H-R diagram by \citet{Meyer_phd}
using models of \citet{DM94}. The reddening vector ($A_{V} = 10$ mag) is also shown,
\citep{Cohen} as is the 90 \% completeness threshold discussed above.
The subsample used to calculate the ratio is extinction limited at $A_{V} = 18$ mag
and $A_{V} = 17.1$ mag for the 3 $\times\ 10^{5}$ yr and 1 Myr isochrones, respectively.
The samples were extinction limited to ensure that we were sensitive to all
masses from 0.1 to $10.0 M_{\sun}$ uniformly, and not sampling massive, more
luminous objects deeper into the cloud \citep{Meyer_phd}.  The number of objects
used in calculating the ratio for each model set ranged from 12 to 19.

Masses of objects were determined by dereddening onto an isochrone, assuming
no infrared excess in the J and H-bands.  In order to evaluate the possible
effect of infrared excess, a (J - H) vs. (H - K) color-color diagram was created
for 67 of the sources using K-band photometry from \citet{Meyer_phd}
matched to sources detected in this study (Figure \ref{fig:colcol}).  K-band photometry
for close companions unresolved in \citet{Meyer_phd} was taken from BSC03.
The JHK color-color plot shows that 46 out of 67 sources are reddened main sequence
(RMS) or weak-lined T Tauri stars (WTTS).  Fourteen sources are classical T Tauri
stars (CTTS), and seven are objects with more extreme infrared colors.
In our extinction limited sample, $6 / 16 = 38 \pm \ 15$ of the sources have IR
excess, which is similar to the fraction of stars with IR excess found by
\citet{Meyer_phd} and \citet{hai00}.
Even with the significant fraction of IR excess objects, we do not expect them
to affect our results greatly.  The expected J and H-band excess for a typical CTTS
would result in an effect only on the order of our photometric errors
\citep{Meyer97}.  If the infrared excess was greater than expected, as is possible
for the 7 more extreme IR-excess sources, one would expect the excess to be
greater in H-band than J-band.  This would lead to an over-estimation of an
object's reddening and luminosity, and therefore mass.  Hence the ratio of
intermediate to low mass objects would be over-estimated as well.  However,
the three extreme objects in the extinction limited sample have already been
classified as low mass objects, therefore we are confident that these
objects have been correctly characterized as low mass objects.

The ratio of intermediate to low mass objects was determined
to be $0.32 \pm\ 0.22$, consistent with the expected ratio using the
IMF presented in \citet{MS79} and \citet{K01} (see Section \ref{disc_imf} below
for further discussion).  We also determined this ratio for several different
isochrones in order to ascertain the effects of 1) the uncertainty in cluster age,
2) different PMS models, and 3) our color transformations.
Isochrones used in this analysis included the 3 $\times\ 10^{5}$ yr and
1 Myr DM98 models in the CIT magnitude system and the 1 Myr
(youngest available) \citet{BCAH98} isochrones in both the CIT and natural
NICMOS magnitude systems.  For each isochrone, the ratio of intermediate
to low mass objects was determined.
These results are listed in Table \ref{tab:CMDresults} and discussed below.

Figure \ref{fig:cmdresult} shows a distribution of expected ratios of
intermediate (1.0 to $10 M_{\sun}$) to low (0.1 to $1.0 M_{\sun}$) mass objects
for a Miller-Scalo IMF \citep{MS79} along with
the ratios determined from the cluster CMD using the different isochrones.
Ten thousand artificial
samples of 20 stars were generated assuming a Miller-Scalo IMF and the
relative probabilities of obtaining different ratios of intermediate to low
mass stars are indicated by the vertical bars.
The data points show the derived ratios for each set of models in each magnitude
system.  Errors shown are $\sqrt{n}$ errors from counting the number of
intermediate and low mass objects.  Using the DM98 models and
assuming a cluster age of 3 $\times\ 10^{5}$ yr yielded good agreement between
the cluster ratio and the ratio expected in a field star IMF.  Adopting an
older cluster age resulted in a higher fraction of intermediate mass objects in
all cases, due to the luminosity evolution as a function of time for a given
mass star.  The \citet{BCAH98} isochrones result in a higher ratio
of intermediate mass objects than do the DM98 isochrones.  It is, however,
important to note that the transformation of the pre-main sequence tracks to the
observational plane were done in a different manner for the two sets of
models.  \citet{BCAH98} derived colors and magnitudes by convolving their
evolutionary models with stellar atmosphere models in the CIT and NICMOS systems.
The DM98 tracks were transformed to the observational plane by adopting bolometric
corrections and intrinsic colors of dwarf stars \citep[Appendix C]{Meyer_phd}.
This difference in transformation between theoretical and observational planes
may be reflected in the difference in the ratios calculated for the two sets of
models. The bottom two data points show the difference in the derived
ratio of intermediate to low mass objects between the CIT
and NICMOS magnitude systems.  These points indicate that the untransformed
magnitudes in the natural NICMOS system yield a lower ratio.  Again, this
difference may be a result of different methods by which the models were transformed
to the observational plane, or our color transformations, or both.  Our results
indicate that the ratio of intermediate to low mass objects in NGC 2024 is
consistent with the field star IMF in the solar neighborhood for all of the
models used, as well as for the CIT and NICMOS magnitude systems.

The measured ratio of intermediate to low mass objects was also compared to
a theoretical value expected from the field IMF presented by \citet{K01}.  By
integrating the IMF over the two appropriate mass ranges and taking the ratio
of the integrals, we find a theoretical ratio of $0.199 \pm 0.103$.  This
$1 \sigma$ error was calculated for the binomial
distribution (with a sample of 15 stars and a probability of success taken
as $p = 0.199$, the probability of choosing an intermediate mass star).  We find
that our results for NGC 2024 are consistent with the ratio expected from a
\citet{K01} IMF.

\subsection{Discussion of the CMD and IMF} \label{disc_imf}

Previous studies have investigated the nature of the IMF, both in the field
and in star-forming regions.  Our results have found that none of the models
used, or ages assumed, result in a ratio of intermediate to low mass objects
that is inconsistent with the field star IMF determined by \citet{MS79} or
\citet{K01}.  \citet{CRR96} found the power law slope of the IMF in NGC 2024
to be $-1.2$ in linear mass units for low mass ($0.04 \lesssim M/M_{\sun}
\lesssim 0.5$) objects, similar to the form
of the IMF derived in \citet{K01} for field stars.  Our
results appear to confirm the agreement between the IMF in NGC 2024 and the
field for stellar objects within the mass range of $0.1 M_{\sun}$ to
$10.0 M_{\sun}$.  It should be noted, however, that because our sample includes
a small number of stars in the extinction limited sample, that the IMF of the
region would need to differ very significantly from the field star IMF for any
effect to be detectable.

Our results also show that color transformations and
the different models and ages used result in small differences in the ratio of
intermediate to low mass objects, but that the values agree within errors.
We make no conclusions regarding the relative abundance of brown
dwarfs, as our study is not sensitive to a significant number of objects with masses in
the brown dwarf regime.  Within the Av-limited sample, the least massive object in
our study had a mass at the hydrogen burning limit.  The substellar IMF for the Trapezium
\citep{Luhman} and IC 348 \citep{NTC00} have been recently investigated, and if the
IMF in star-forming regions is indeed very similar, as current evidence suggests, one
may expect to find a significant number of substellar objects in NGC 2024 as well.

\section{The Binary Fraction} \label{sec-bf}
\subsection{Binary Fraction Results} \label{sec-binfrac}

Each of the 95 sources detected in both bands was carefully inspected for evidence
of multiplicity.  Three binary candidates were identified in the sample with
separations between $0\arcsec .4$ and $1\arcsec .0$ and up to a difference of 4 mag between
the primary and companion.  All companions were verified in both the J- and
H-bands. This angular separation range translates into a
projected linear separation range of 184 to 460 AU assuming a cluster distance
of 460 pc.  We assume that the statistical distribution of true semi-major axes
is the same as the distribution of observed separations \citep[and references
therein]{RZ93}.  The outer limit of 1".0 was set to ensure that the probability of
a coincidental superposition of stars, given the surface density of sources in
the cluster, did not exceed 1 \%\footnote[1]{We examined the number of companions out to
$2\arcsec$ and found the number of additional associations expected from chance
projections.}.  Given the average surface density
of the cluster and the 95 detected sources, it is possible that one of these
candidates is a coincidental association.  A difference of 4 mag between
the primary and companion corresponds to a mass ratio of $q = M_{comp} / M_{pri}
\approx 0.1$ at an age of 3 $\times 10^{5}$ yr.  As the 95 potential primaries ranged
in mass from $\sim 0.1 M_{\sun}$ to $2.3 M_{\sun}$, our characteristic primary
mass was taken to be $\sim 1.0 M_{\sun}$.  The binary candidates, their separations
from their primaries, and $\Delta m_{J}$ are listed in Table \ref{tab:binaries}.

Since these binary candidates were intially detected by \emph{daofind} as a single
source, photometry was performed separately on each of the companions by
visually determining the centroid of each companion, and using a smaller
aperture to perform the photometry (aperture of 2 pixels with a sky annulus of
5-7 pixels).  The appropriate aperture corrections were made for the new
aperture size.  The mass of each object was determined by dereddening onto the
$3 \times\ 10^{5}$ yr DM98 isochrone, and values of q were calculated.  The q
values for each binary are listed in Table \ref{tab:binaries}

A series of artificial star tests were performed in order to assess the competeness
of our visual detection technique in identifying companions over the separations
and magnitude differences described above.  Artificial stars were randomly generated
with a range of separations and $\Delta m_{J}$ and placed as companions to real
sources in several of the data frames.  The artificial companions were then subject
to the same visual inspection used to find the binary candidates.
Table \ref{tab:bincomp} shows the recovery fraction of artificial companions as a
function of separation and difference in magnitude between the primary and
companion.  The number of artificially generated companions in each
separation-magnitude bin was between 10 and 20 with a median of 11.  None of the
artificial stars generated were below the 90 \% completeness limits
for the photometry described in Section \ref{sec-magcal}.

Binary detections were 90 \% complete for the following range of separations
and mass ratios.  At 184 AU (2 pixels) we were sensitive to equal mass companions
($q = 1$).  At 276 AU, $q = 0.2$ was the detection limit, and
from ~325 AU to 460 AU detections were sensitive down to $q = 0.1$.  Over this
range of separations and mass ratios, 3 binaries out of 95 potential primaries
yielded a binary fraction of $0.032^{+0.018}_{-0.025}$.  This error includes $\sqrt{n}$
counting errors and the possibility of a coincidental superposition of stars.

It is interesting to compare the sample of binaries detected in this study
to that of BSC03.  Out of our sample of three binaries, one of them, 4-11a \& b
is also detected by BSC03.  The derived separations are comparable, $0\arcsec.42$
in this study and $0\arcsec.39$ in the BSC03 study.  The other binaries found in
this study, 1-11a \& b and 3-5a \& b, were not resolved by BSC03, as they were not
observed with AO by their study.  There were also two binaries identified BSC03 not
identified in this study.  The first (BSC03 ID 88 \& 89) only detected as a single
source (ID 2-7) by this study, with no evidence of a companion.  It is possible that
the companion to this star is deeply embedded, hence only detected in the K-band.
The other binary (BSC03 ID 54 \& 55), source 7-5 in this study, appears to be marginally
resolved in the J-band, with some evidence of multiplicity and a separation of
$\lesssim 0\arcsec.4$ in our J-band image. However, the detection was unverified in
the H-band, as the brightness of the object makes it difficult to resolve
close companions.  Additionally, the close separation
suggests that it would have fallen out of the separation range probed by this study.
Three of the pairs identified by BSC03 (ID 67 \& 64; 70 \& 71; 70 \& 72) were
detected by this study, but have separations greater than the maximum separation
considered.  The other binaries detected by BSC03 not mentioned above fall outside
of the field-of-view of our observations.

Our desire was to compare the binary fraction of NGC 2024 to that of solar
neighborhood field stars determined by \citet[hereafter DM91]{DM91}.  Thus it
was necessary to be certain that we were comparing equivalent separations and
mass ratios between the two studies.  Our study was sensitive to companions
with linear projected separations of 184 to 460 AU, corresponding to an orbital
period of log P [days]$\ \approx 6.0$ to 6.5 for a characteristic system mass of
$1.0 M_{\sun}$. This is approximately one-half bin in
Figure 7 of DM91 which shows the companion period distribution in their study.
The following calculations were made to arrive at an expected number of binaries
from the companion distribution in DM91 for each of our four separation bins
(centered upon 184, 276, 368, and 460 AU).  In the range
log P [days]$\ = 6.0$ to 7.0, DM91 found a binary fraction of 8.5 \%.  Our separation
range covers only half the DM91 bin, so the binary fraction for 184 to 460 AU was
taken to be 4.25 \%.  There were 95 potential primaries detected in our sample,
leading to a expected number of binaries to be $95 \times 0.0425 = 4$ binaries,
assumed to be evenly distributed over our separation range.
However, this assumes that we were sensitive to the same range of companion masses
as DM91 over the entire separation range, which was not the case.  To correct for
this affect, we attempted to account for differences in the range of
companion masses.   For the mass of a typical primary in our sample
($1.0 M_{\sun}$), we performed an integral of the companion mass
distribution (taking the distribution to be a single power law :$\frac{dN}{dM}
\sim M^{-\alpha}$) over the mass range to which we were 90 \% complete.
This was compared to the number of expected companions down to $q = 0.1$.  The
ratio between the two integrals was used to adjust the expected number of
binaries in each separation bin, to account for the fact that our mass
sensitivity range was smaller than that of DM91.  Correcting for this difference
in sensitivity to low mass companions at close separations yielded 2.9
expected binaries over our separation range.  Varying the power law index
of the companion mass distribution between 1.5 and 0.5 yielded between 2.7 and 3.0
binaries over our separation range.  One may note that this correction for mass
sensitivity only reduced the expected number of binaries by about 1, which is
within the errors for the number of detected binaries given the small number of
detected binaries (i.e., the number of expected binaries both before and after the
correction is consistent with the observed number).  However, the correction
does result in a decrease of about 25\% in the expected number of binaries, hence for
the most accurate possible comparison between the two studies, this step is crucial.
Additionally, any application of this type of mass correction to larger samples would
result in a more significant effect relative to counting errors.

Our study detected $3^{+1.8}_{-2.5}$
binaries, consistent with the expected number for the solar neighborhood fraction.
Table \ref{tab:expectbin} summarizes the expected and detected number of binaries
in each separation bins assuming different functional forms of the companion
mass distribution.  It is interesting to note that all of our detected binaries
were at close separations, while we would expect to observe more binaries in the
outer separations.  This is consistent with the findings of \citet{RZ93} who
found the distribution of companion separations (for nearby southern dark clouds)
to be a steeply rising function toward smaller separations.

The binary fraction of star-forming regions in the Orion B cloud have been the
focus of previous studies, including \citet{PSG97}, which presented
HST WFPC2 observations of three clusters (NGC 2024, NGC 2068, and NGC 2071). They
found a binary fraction of $0.15 \pm 0.04$ for a separation range of 138 to 1050 AU,
which was about 1.3 times greater than solar neighborhood G and K stars.
This suggests a slight excess of binaries in these regions;
however, the signifigance of the excess with respect to the field was not high
\citep{PSG97}.  Our results are also in agreement with BSC03 who
determined the binary fraction of NGC 2024 to be similar to that of the solar
neighborhood using K-band adaptive optics observations.

\subsection{Discussion of the Binary Fraction}

Does the local environment of a star-forming region (characteristics such as stellar
density) correlate with its binary fraction?  We might expect that regions of
higher stellar density are subject to more frequent interactions, which would disrupt
binaries and lead to a lower binary fraction.  To address this
question we compared our results for the binary fraction of NGC 2024 to those
of other star-forming regions in order to determine how NGC 2024 fits into the
context of galactic star formation.  Specifically, we evaluate whether a correlation
can be found between the binary fraction of a young cluster and its central stellar
density.

We begin by considering the binary fraction of NGC 2024 compared to star-forming
regions of lower density.  Two well-studied regions which fall into this category
are the Taurus star-forming region and the Rho Ophiuchus dark cloud.  The stellar
density of $\rho \ $ Ophiuchus was determined to be more than a factor of two smaller
than that of NGC 2024 (4000 $pc^{-3}$).  The density of Taurus is smaller still
\citep{Lada91,Lada93}. \citet[hereafter GNM]{GNM93} performed K-band
speckle imaging of the Taurus-Auriga and Ophiuchus-Scorpius star-forming
regions and found a binary fraction of $60 \pm 17$ \%, about 4 times greater
than DM91 for the same separation range.  It is difficult to compare the binary
fraction in the same \emph{physical separation range} between these studies and
ours, as the outermost separation probed by GNM is 252 AU,
only 68 AU greater than the innermost separation for this study.  However,
in the overlap region of 184 to 252 AU, the binary fractions are
2/24 = $8.3 \pm 5.9$\% and 1/45 = $2.2 \pm 2.2$\% for Tau-Aur and Oph-Sco,
respectively.  In the overlap region, NGC 2024 has two binaries, and a binary
fraction of 2/95 = $2.1 \pm 1.5$\%.  While it is impossible to make a conclusive
comparison due to the small sample size, the results do seem to suggest that the
binary fraction of NGC 2024 is lower than that of Taurus-Auriga.
\citet{Leinert} observed the Taurus-Auriga region
with K-band speckle imaging and found a binary fration of $42 \pm 6$ \%,
again enhanced relative to the field star fraction of DM91.  Their angular
separation range of 0".13 to 13" translates to a linear projected distance
of 18 to 349 AU assuming a distance of 140 pc. For the same linear separation
range, DM91 found a binary fraction of about 20 \%.
Three low density star-forming regions in the southern sky are found in the dark
clouds Chameleon, Lupus, and Corona Australis.  \citet{Ghez97} conducted a survey of
these regions and found the binary fraction to be similar to Taurus and Ophiuchus,
enhanced relative to the solar neighborhood (However, see \citet{Kohler01}). These
results demonstrate that low density star-forming environments like those in
Taurus seem to have significantly larger binary fractions than the environments
in Orion and the field.  This may indicate that the primary mode of star
formation in the galaxy is not through T associations like Taurus.

The Trapezium cluster in Orion is a region of very high stellar density, about 3.5
times greater than that of NGC 2024 \citep{Lada91}.
The adaptive optics study of the Trapezium by \citet{SCB99} determined the
binary fraction to be the same as DM91 for a separation range of 132 to 264 AU.
These results were also similar to a
study by \citet{Petr}, who determined the binary fraction in the central 40"
$\times$ 40" of the cluster using speckle observations in the $K_{s}$ and H bands.
They found that the binary fraction agreed with that of main-sequence field
stars, and a factor of 3 times lower than the Taurus-Auriga star-forming region.
We found that our binary fraction for NGC 2024 was roughly equivalent
to that of solar neighborhood field stars over the same separation
range, after being corrected for differences in mass sensitivity.
Thus, with regard to the binary fraction, it appears that the intermediate density
young cluster in NGC 2024 is more similar to very dense clusters like Trapezium
than low density star-forming regions such as Taurus.
The similarity of the binary fraction in these dense regions compared to the field
suggests the possibility that most galactic field stars tend to form in denser,
richer environments.  However, there is some evidence to suggest that Trapezium will
form a bound cluster \citep{kah01,HH98} and \citet{am01} estimates only a relatively
small percentage of stars ($\sim 10\%$) in the galaxy form in regions which
evolve into bound clusters.  This may indicate that galactic field stars
form in regions of relatively high stellar density, like NGC 2024, but not so high
as to form bound clusters.

Figure \ref{fig:bf_density} shows the binary fraction relative to the field versus
the stellar density for five different star-forming regions with ages estimated to
be 2 Myr or less.  It shows that young clusters of high or intermediate densities
have binary fractions similar to the field while Taurus, significantly
less dense than the other regions, has a binary fraction enhanced
relative to the field.  A linear regression to the data points yields a fit with
a decreasing binary fraction with increasing density.  The correlation
coefficient for the fit implies a $\sim 80 \%$ probability that the
fitted trend is not the result of a random distribution of points, which suggests
that the trend is real.  While the cause of the low binary fraction
in regions of higher stellar density
is not known, one possibility is that binaries are disrupted by dynamical
interactions.  Following the argument in \citet{bt}, the lifetime of
a soft binary ($\vert E \vert < m \sigma^{2}$), the category in which all of the
binaries in this study fall, is on the order of several Myr.
As this relation varies inversely with the density of a
given star forming region, one would expect to see a decrease in binary fraction with
increasing stellar density in the regime where disruptions of binaries is ongoing,
that is the age of the cluster is less than the lifetime of a typical soft binary.
Although the number of clusters plotted is small, and the errors fairly large, there
does seem to be a trend consistent with this line of reasoning.

\section{Summary and Conclusions} \label{sec-conc}

We have presented high resolution images of the embedded cluster in NGC 2024
taken with NICMOS onboard \emph{HST}.  Ninety-five sources were detected in the F110W
and F160W bands, and photometry was extracted for 79 of these sources and transformed
into $m_{J}$ and $m_{H}$ magnitudes in the CIT system.  A color-magnitude diagram
was plotted, and the masses for the sources were determined using the pre-main
sequence stellar models of \citet{DM98} and \citet{BCAH98}.  The ratio of
intermediate ($1.0 - 10.0 M_{\sun}$) to low ($0.1 - 1.0 M_{\sun}$) mass objects was
found to be consistent with a field star IMF for isochrones of $3 \times 10^{5}$ yr
and 1 Myr.

A study of the binary fraction of the cluster was also completed.  Three binary
candidates were detected out of 95 potential primaries, resulting in a binary
fraction of $3.2^{+1.8}_{-2.5}$ \% for a linear projected separation range of
184 to 460 AU.  The completeness region in separation - companion mass space was
determined using a series of artificial star tests and showed the study to be
complete down to $q = 0.1$ outside of 368 AU, assuming a typical primary mass of
$1.0\ M_{\sun}$.  The companion mass sensitivity was reduced for closer separations.
At the innermost separation of 184 AU, we were sensitive to equal mass
companions.  When our binary sample was corrected for differences in companion
mass sensitivity, we found the binary fraction of NGC 2024 to be consistent with
the solar neighborhood binary fraction determined by \citet{DM91} and \citet{K01}.
A comparison to \citet{GNM93} suggests that the binary fraction of NGC 2024 is
lower than that of the low density star-forming regions in Taurus-Auriga.  This
result supports the notion that the much of the star formation in the galaxy occurs
in intermediate density clusters such as NGC 2024 and not within loose T associations
like those found in Taurus.

\section{Acknowledgments}
WL was supported by NASA under the NICMOS grant to the University of Arizona.
MRM is very grateful for support from the Lucas Foundation, and NASA grants
HF-01098.01-97A and GF-7417 awarded by the Space Telescope Science Institute
which is operated by the Association of Universities for Research in Astronomy,
Inc. for NASA under contract NAS 5-26555.  The authors also thank the referee for
helpful comments.

{}

\clearpage
\begin{figure}
\caption{Two-color (F110W and F160W) image of the young embedded cluster
in NGC 2024.  The field-of-view is just over $2\arcmin\ \times\ 2\arcmin$.  North is up and east is
to the left.}
\label{fig:ngc2024}
\end{figure}

\begin{figure}
\caption{Distribution of souces detected in the J and H bands.  Frame numbers are
shown in the
upper-right corner of each frame.  Plate scale is $\approx$ 0.2 arcsec per pixel
and each frame overlaps adjacent frames by $\approx$ 40 pixels.  The center of
Frame 1 is located at 05h 41m 42.4s, -1$\degr$ $55\arcmin$ 15$\arcsec$.9}
\label{fig:star_pos}
\end{figure}

\begin{figure}
\caption{The CMD of the cluster in the CIT magnitude system.
Also shown are the 300,000 yr (DM98) isochrone,
0.1 $M_{\sun}$ extinction limit, and 90 \% completeness limit.}
\label{fig:cmd}
\end{figure}

\begin{figure}
\caption{Color-color diagram for 67 objects in the sample.  Forty-six
are reddened main sequence (RMS) or weak-lined T Tauri stars (WTTS), 14 are
classical T Tauri stars (CTTS) and 7 are objects with more extreme colors. K-band
data was taken from \citet{Meyer_phd} and \citet{bsc}.  The CTTS locus
is that of \citet{Meyer_phd} and the stellar models are those of \citet{DM94}}
\label{fig:colcol}
\end{figure}

\begin{figure}
\caption{The ratio of intermediate to low mass objects in the cluster as determined
by this study (data points and errors), and a distribution of ratios predicted by a
Miller -Scalo IMF (MS79) for 20 objects (vertical bars).  The vertical spacing
of the data points is arbitrary.}
\label{fig:cmdresult}
\end{figure}

\begin{figure}
\caption{The binary fraction relative to the field vs. stellar density for
five young star-forming regions.  Density data was taken from Table 1
of \citet{Meyer_ppiv} and binary fractions are those compiled in Table 1 of
\citet{Mathieu}, except for NGC 2024 which reflects the value found in this
study.}
\label{fig:bf_density}
\end{figure}

\clearpage

\begin{deluxetable}{ccccccccccc}
\tablecaption{Objects Dectected in J and H bands \label{tab:objects}}
\tablewidth{0pt}
\rotate
\tablehead{
\colhead{Frame-ID} &
\colhead{X} &
\colhead{Y} &
\colhead{RA (J2000)} &
\colhead{Dec (J2000)} &
\colhead{J} &
\colhead{J-error} &
\colhead{H} &
\colhead{H-error} &
\colhead{BSC03 ID} &
\colhead{Notes}
}
\startdata
1-1 & 71.0 & 21.0 & 05 41 41.30 & -1 53 51.9 & 17.42 & 0.35 & 13.62 & 0.14 & 66 & proximity to FIR2\\	
1-2 & 233.0 & 29.0 & 5 41 43.49 & -1 53 56.3 & 18.85 & 0.59 & 14.85 & 0.18& N/A\\	
1-3 & 200.8 & 34.6 & 5 41 43.04 & -1 53 56.9 & 18.81 & 0.58 & 15.00 & 0.18 & 79\\
1-4 & 161.0 & 66.0 & 5 41 42.47 & -1 54 2.6 & 16.81 & 0.29 & 13.55 & 0.14 & 75\\	
1-5 & 160.1 & 99.9 & 5 41 42.42 & -1 54 9.4 & 15.11 & 0.20 & 12.51 & 0.13 & 74 & proximity to FIR3\\	
1-6 & 43.1 & 104.4 & 5 41 40.82 & -1 54 8.3 & 20.74 & 1.32 & 16.01 & 0.23 & N/A\\	
1-7 & 98.5 & 117.5 & 5 41 41.56 & -1 54 11.9 & 17.69 & 0.38 & 14.04 & 0.15 & 70\\
1-8 & 107.0 & 118.0 & 5 41 41.68 & -1 54 12.2 & 18.05 & 0.44 & 14.84 & 0.18 & 72\\	
1-9 & 214.9 & 120.3 & 5 41 43.14 & -1 54 14.5 & 19.05 & 0.69 & 16.29 & 0.25 & 81\\	
1-10 & 102.8 & 124.7 & 5 41 41.61 & -1 54 13.5 & 17.52 & 0.36 & 14.19 & 0.16 & 71\\	
1-11a & 128.6 & 172.8 & 5 41 41.91 & -1 54 23.7 & 16.77 & 0.22 & 12.98 & 0.13 & 73\\
1-11b & 130.0 & 176.0 & 5 41 41.92 & -1 54 24.4 & 18.67 & 0.36 & 15.10 & 0.16 & 73\\
1-12 & 225.6 & 215.2 & 5 41 43.17 & -1 54 34.0 & 17.01 & 0.32 & 14.42 & 0.16 & 82 & proximity to FIR4\\	
1-13 & 194.3 & 227.4 & 5 41 42.74 & -1 54 35.9 & 14.71 & 0.18 & 12.06 & 0.12 & 78\\
1-14 & 77.6 & 48.0 & 5 41 41.36 & -1 53 57.5 & 21.30 & 1.77 & 17.34 & 0.34 & 68\\
2-1 & 147.0 & 17.0 & 5 41 45.33 & -1 53 56.2 & N/A & N/A & N/A & N/A & N/A\\	
2-2 & 65.9 & 68.0 & 5 41 44.17 & -1 54 5.2 & 18.56 & 0.51 & 14.26 & 0.16 & N/A\\
2-3 & 127.7 & 68.7 & 5 41 45.01 & -1 54 6.4 & 17.60 & 0.36 & 12.94 & 0.13 & N/A\\	
2-4 & 191.1 & 86.4 & 5 41 45.85 & -1 54 11.1 & N/A & N/A & 13.02 & 0.13 & N/A\\
2-5 & 209.0 & 103.9 & 5 41 46.07 & -1 54 14.9 & N/A & N/A & 14.56 & 0.17 & 93\\	
2-6 & 82.6 & 141.4 & 5 41 44.32 & -1 54 20.3 & 17.21 & 0.33 & 13.75 & 0.15 & 84\\	
2-7 & 147.6 & 148.9 & 5 41 45.19 & -1 54 23.0 & 17.14 & 0.32 & 13.39 & 0.14 & 88,89\\	
2-8 & 168.0 & 175.5 & 5 41 45.44 & -1 54 28.7 & N/A & N/A & N/A & N/A & 90\\
2-9 & 123.8 & 214.9 & 5 41 44.79 & -1 54 35.9 & 15.99 & 0.24 & 12.71 & 0.13 & 86\\	
2-10 & 83.1 & 26.8 & 5 41 44.45 & -1 53 57.1 & 21.70 & 2.19 & 17.02 & 0.31 & N/A\\
2-11 & 118.0 & 162.0 & 5 41 44.77 & -1 54 25.1 & 12.97 & 0.15 & 10.61 & 0.11 & 85\\	
2-12 & 192.0 & 179.0 & 5 41 45.76 & -1 54 29.8 & N/A & N/A & N/A & N/A & 92 & IRS2\\
3-1 & 151.2 & 47.0 & 5 41 45.10 & -1 54 47.3 & 15.05 & 0.19 & 11.88 & 0.12 & 87\\	
3-2 & 148.8 & 85.3 & 5 41 45.02 & -1 54 55.0 & 18.76 & 0.56 & 14.69 & 0.17 & N/A\\	
3-3 & 90.6 & 93.0 & 5 41 44.23 & -1 54 55.6 & 19.62 & 0.80 & 15.31 & 0.19 & N/A\\	
3-4 & 200.6 & 96.9 & 5 41 45.71 & -1 54 58.3 & 18.65 & 0.61 & 16.23 & 0.24 & N/A\\	
3-5a & 219.0 & 114.0 & 5 41 45.94 & -1 55 2.1 & 16.66 & 0.29 & 13.23 & 0.14 & N/A\\
3-5b & 220.0 & 115.0 & 5 41 45.96 & -1 55 2.4 & 16.58 & 0.28 & 13.34 & 0.14 & N/A\\	
3-6 & 172.8 & 122.6 & 5 41 45.31 & -1 55 3.0 & 17.36 & 0.35 & 13.80 & 0.15 & N/A\\	
3-7 & 201.0 & 131.0 & 5 41 45.68 & -1 55 5.2 & 17.36 & 0.34 & 13.85 & 0.16 & N/A\\
3-8 & 79.7 & 154.2 & 5 41 44.01 & -1 55 7.8 & 18.81 & 0.59 & 15.32 & 0.19 & N/A\\	
3-9 & 96.3 & 216.6 & 5 41 44.16 & -1 55 20.8 & 18.58 & 0.52 & 14.44 & 0.16 & N/A\\	
3-10 & 113.2 & 228.0 & 5 41 44.38 & -1 55 23.4 & 13.44 & 0.16 & 11.25 & 0.11 & N/A\\	
3-11 & 109.0 & 237.1 & 5 41 44.31 & -1 55 25.1 & 14.65 & 0.19 & 12.94 & 0.13 & N/A\\	
4-1 & 198.1 & 24.9 & 5 41 42.77 & -1 54 39.8 & 17.93 & 0.45 & 15.43 & 0.20 & 77\\	
4-2 & 104.1 & 32.9 & 5 41 41.48 & -1 54 39.8 & 15.96 & 0.23 & 12.81 & 0.13 & 67\\	
4-3 & 117.2 & 37.2 & 5 41 41.65 & -1 54 40.9 & 19.98 & 1.02 & 16.17 & 0.24 & 69\\	
4-4 & 99.0 & 59.3 & 5 41 41.38 & -1 54 45.1 & 17.75 & 0.39 & 13.69 &0.15 & 65\\	
4-5 & 135.9 & 132.9 & 5 41 41.80 & -1 55 0.6 & 17.21 & 0.33 & 13.99 & 0.15 & N/A\\	
4-6 & 185.0 & 141.2 & 5 41 42.45 & -1 55 3.2 & 15.94 & 0.23 & 12.68 & 0.13 & N/A\\	
4-7 & 169.3 & 179.0 & 5 41 42.20 & -1 55 10.6 & 12.74 & 0.15 & 11.20 & 0.11 & N/A\\
4-8 & 55.6 & 231.9 & 5 41 40.59 & -1 55 19.3 & 15.70 & 0.22 & 12.83 & 0.13 & N/A\\	
4-9 & 38.9 & 251.7 & 5 41 40.35 & -1 55 23.1 & 21.43 & 2.21 & 16.76 & 0.29 & N/A\\	
4-10 & 190.0 & 57.1 & 5 41 42.62 & -1 54 46.2 & 14.80 & 0.19 & 12.380 & 0.12 & N/A\\	
4-11a & 95.3 & 25.0 & 5 41 41.37 & -1 54 38.1 & 18.30 & 0.51 & 15.11 & 0.19 & 63\\	
4-11b & 93.8 & 26.5 & 5 41 41.35 & -1 54 38.4 & 18.50 & 0.55 & 15.32 & 0.20 & 64\\	
5-1 & 187.4 & 57.3 & 5 41 39.59 & -1 54 42.4 & 18.15 & 0.46 & 14.59 & 0.17 & 57\\	
5-2 & 91.9 & 131.4 & 5 41 38.20 & -1 54 55.8 & 14.90 & 0.19 & 12.80 & 0.13 & 40\\	
5-3 & 158.9 & 155.0 & 5 41 39.09 & -1 55 1.7 & 17.20 & 0.34 & 14.65 & 0.17 & N/A\\	
5-4 & 169.6 & 179.3 & 5 41 39.20 & -1 55 6.8 & 19.79 & 1.25 & 18.54 & 0.55 & N/A\\	
5-5 & 65.6 & 207.0 & 5 41 37.76 & -1 55 10.6 & 16.32 & 0.27 & 14.67 & 0.17 & 31\\
5-6 & 53.1 & 66.0 & 5 41 37.75 & -1 54 41.9 & 13.25 & 0.15 & 11.79 & 0.12 & 32\\	
5-7 & 58.5 & 42.5 & 5 41 37.85 & -1 54 37.2 & N/A & N/A & N/A & N/A & 33 & IRS1\\
6-1 & 32.7 & 45.2 & 5 41 37.75 & -1 53 52.3 & N/A & N/A & N/A & N/A & 30\\
6-2 & 60.0 & 71.3 & 5 41 38.09 & -1 53 58.1 & 13.34 & 0.15 & 11.54 & 0.12 & 37\\	
6-3 & 135.0 & 70.6 & 5 41 39.11 & -1 53 59.2 & N/A & N/A & N/A & N/A & 50\\	
6-4 & 148.7 & 88.0 & 5 41 39.28 & -1 54 3.0 & 16.51 & 0.28 & 14.59 & 0.17 & 52\\	
6-5 & 220.8 & 108.1 & 5 41 40.23 & -1 54 8.3 & N/A & N/A & 14.11 & 0.15 & 60\\	
6-6 & 149.3 & 147.7 & 5 41 39.22 & -1 54 15.1 & 14.01 & 0.17 & 11.12 & 0.11 & 51\\	
6-7 & 79.8 & 170.7 & 5 41 38.25 & -1 54 18.6 & 17.43 & 0.40 & 15.51 & 0.20 & 42\\	
6-8 & 101.0 & 207.1 & 5 41 38.50 & -1 54 26.3 & 11.01 & 0.13 & 11.28 & 0.11 & 46\\	
6-9 & 121.1 & 224.4 & 5 41 38.75 & -1 54 30.2 & 17.03 & 0.32 & 14.34 & 0.16 & 48\\	
6-10 & 39.0 & 209.1 & 5 41 37.65 & -1 54 25.7 & 12.49 & 0.14 & 11.09 & 0.11 & 28\\	
7-1 & 123.6 & 44.0 & 5 41 39.24 & -1 53 8.7 & 17.98 & 0.43 & 14.63 & 0.17 & 53\\	
7-2 & 135.4 & 70.8 & 5 41 39.37 & -1 53 14.4 & 19.16 & 0.86 & 17.89 & 0.41 & N/A\\	
7-3 & 84.0 & 122.7 & 5 41 38.62 & -1 53 24.0 & N/A & N/A & N/A & N/A & 47\\
7-4 & 225.0 & 133.0 & 5 41 40.52 & -1 53 28.5 & N/A & N/A & 15.52 & 0.20 & N/A\\	
7-5 & 152.5 & 136.9 & 5 41 39.53 & -1 53 28.1 & 13.36 & 0.16 & 10.82 & 0.11 & 54,55 & possible close binary\\	
7-6 & 53.0 & 139.0 & 5 41 38.18 & -1 53 26.8 & 16.52 & 0.29 & 14.85 & 0.18 & 39\\	
7-7 & 67.0 & 152.0 & 5 41 38.35 & -1 53 29.6 & 15.93 & 0.24 & 14.08 & 0.15 & 44\\	
7-8 & 207.0 & 168.5 & 5 41 40.23 & -1 53 35.4 & 14.87 & 0.19 & 11.97 & 0.12 & N/A\\	
7-9 & 155.9 & 170.6 & 5 41 39.54 & -1 53 34.9 & 15.18 & 0.20 & 12.56 & 0.13 & 56\\	
7-10 & 69.2 & 175.0 & 5 41 38.36 & -1 53 34.4 & 12.87 & 0.15 & 12.57 & 0.13 & 45\\	
7-11 & 198.2 & 176.1 & 5 41 40.10 & -1 53 36.8 & 15.30 & 0.20 & 12.71 & 0.13 & 58\\	
7-12 & 187.7 & 199.0 & 5 41 39.94 & -1 53 41.3 & 18.03 & 0.48 & 15.95 & 0.23 & N/A\\	
7-13 & 245.5 & 235.4 & 5 41 40.68 & -1 53 49.6 & 18.17 & 0.46 & 14.60 & 0.17 & N/A\\	
7-14 & 53.0 & 57.7 & 5 41 38.27 & -1 53 10.3 & 11.88 & 0.14 & 10.14 & 0.11 & 43\\	
7-15 & 38.0 & 106.7 & 5 41 38.01 & -1 53 20.0 & 15.76 & 0.24 & 14.25 & 0.16 & 36\\	
7-16 & 29.0 & 70.1 & 5 41 37.93 & -1 53 12.4 & 11.28 & 0.13 & 10.41 & 0.11 & 35\\	
8-1 & 197.8 & 73.0 & 5 41 43.21 & -1 53 19.7 & N/A & N/A & N/A & N/A & N/A\\	
8-2 & 216.5 & 129.9 & 5 41 43.40 & -1 53 31.6 & N/A & N/A & 14.89 & 0.18 & N/A\\	
8-3 & 71.4 & 155.1 & 5 41 41.41 & -1 53 34.2 & 19.24 & 0.66 & 14.05 & 0.15 & N/A\\	
8-4 & 53.0 & 160.2 & 5 41 41.15 & -1 53 34.9 & 20.52 & 1.26 & 16.58 & 0.27 & N/A\\	
9-1 & 38.0 & 123.0 & 5 41 44.00 & -1 53 30.9 & N/A & N/A & 13.96 & 0.15 & N/A\\
9-2 & 141.2 & 140.7 & 5 41 45.37 & -1 53 36.3 & 19.28 & 0.74 & 16.26 & 0.24 & N/A\\	
9-3 & 57.6 & 162.3 & 5 41 44.21 & -1 53 39.2 & N/A & N/A & 15.01 & 0.18 & N/A\\
9-4 & 187.1 & 187.7 & 5 41 45.94 & -1 53 46.6 & 16.72 & 0.29 & 14.03 & 0.15 & N/A\\	
9-5 & 214.0 & 194.9 & 5 41 46.29 & -1 53 48.5 & 13.94 & 0.17 & 11.39 & 0.12 & N/A\\	
9-6 & 61.0 & 210.8 & 5 41 44.20 & -1 53 49.1 & N/A & N/A & 15.28 & 0.19 & N/A\\
9-7 & 107.2 & 212.4 & 5 41 44.82 & -1 53 50.2 & N/A & N/A & 15.33 & 0.19 & N/A\\	
9-8 & 47.3 & 231.8 & 5 41 44.00 & -1 53 53.1 & N/A & N/A & 15.02 & 0.18 & N/A\\	
9-9 & 58.449 & 186.4 & 5 41 44.19 & -1 53 44.1 & N/A & N/A & 17.24 & 0.33 & N/A\\	
9-10 & 75.1 & 153.1 & 5 41 44.45 & -1 53 37.6 & N/A & N/A & N/A & N/A & N/A\\
\enddata
\end{deluxetable}

\clearpage

\begin{deluxetable}{ccccc}
\tablecaption{Ratio of Intermediate to Low Mass Objects \label{tab:CMDresults}}
\tablewidth{0pt}
\tablehead{
\colhead{Model Set} &
\colhead{Reference} &
\colhead{Age (Myr)} &
\colhead{Mag. System} &
\colhead{$\frac{N(1 - 10 M_{\sun})}{N(0.1 - 1 M_{\sun})}$}
}
\startdata
1 & \citet{DM98} & 0.3 & CIT & $0.31 \pm 0.22$\\
2 & \citet{DM98} & 1.0 & CIT & $0.67 \pm 0.43$\\
3 & \citet{BCAH98} & 1.0 & CIT & $1.00 \pm 0.63$\\
4 & \citet{BCAH98} & 1.0 & NICMOS & $0.88 \pm 0.64$\\
\enddata
\end{deluxetable}

\clearpage

\begin{deluxetable}{ccccc}
\tablecaption{Binary Candidates Identified \label{tab:binaries}}
\tablewidth{0pt}
\tablehead{
\colhead{ID} &
\colhead{Separation (")/AU} &
\colhead{$m_{J}$ (primary)} &
\colhead{$\Delta m_{J}$} &
\colhead{q = $\frac{M_{comp}}{M_{pri}}$}
}
\startdata
1-11(a,b) & 0.70 / 321 & 16.77 & 1.91 & 0.14\\
3-5(a,b) & 0.36 / 166 & 16.58 & 0.08 & 0.73\\
4-11(a,b) & 0.42 / 195 & 18.30 & 0.20 & 0.92\\
\enddata
\end{deluxetable}

\clearpage

\begin{deluxetable}{ccccc}
\tablecaption{Expected and Detected Binaries \label{tab:expectbin}}
\tablewidth{0pt}
\tablehead{
\colhead{Separation (")/AU} &
\colhead{Detected} &
\colhead{Expected ($\alpha = 0.5$)} &
\colhead{($\alpha = 1.0$)} &
\colhead{($\alpha = 1.5$)}
}
\startdata
0.4 / 184 & 2 & 0.2 & 0.2 & 0.1\\
0.6 / 276 & 1 & 0.8 & 0.7 & 0.6\\
0.8 / 368 & 0 & 1.0 & 1.0 & 1.0\\
1.0 / 460 & 0 & 1.0 & 1.0 & 1.0\\
Total & 3 & 3.0 & 2.9 & 2.7\\
\enddata
\end{deluxetable}

\clearpage

\begin{deluxetable}{cccccc}
\tablecaption{Binary Completeness - Recovery Fraction \label{tab:bincomp}}
\tablewidth{0pt}
\tablehead{
\colhead{$\Delta m_{J}$} &
\colhead{1 pix / 92 AU} &
\colhead{2 / 184} &
\colhead{3 / 276} &
\colhead{4 / 368} &
\colhead{5/ 460}
}
\startdata
0 & 0.000 & 0.909 & 1.000 & 1.000 & 1.000\\
1 & 0.000 & 0.727 & 0.950 & 1.000 & 1.000\\
2 & 0.000 & 0.364 & 0.909 & 1.000 & 1.000\\
3 & 0.000 & 0.000 & 0.500 & 1.000 & 1.000\\
4 & 0.000 & 0.000 & 0.200 & 0.583 & 0.833\\
5 & 0.000 & 0.000 & 0.000 & 0.091 & 0.250\\
\enddata
\end{deluxetable}

\clearpage


\clearpage
\begin{thebibliography}{}

\bibitem[Adams \& Myers(2001)]{am01} Adams, F.~C.~\& Myers,
P.~C.\ 2001, \apj, 553, 744

\bibitem[Baraffe et al.(1998)]{BCAH98} Baraffe, I., Chabrier, G., Allard,
F., \& Hauschildt, P.~H.\ 1998, \aap, 337, 403

\bibitem[Beck, Simon, \& Close(2003)]{bsc} Beck, T.~L.,
Simon, M., \& Close, L.~M.\ 2003, \apj, 583, 358

\bibitem[Binney \& Tremaine(1987)]{bt} Binney, J.~\&
Tremaine, S.\ 1987, Princeton, NJ, Princeton University Press, 1987, 747
p.

\bibitem[Brandner et al.(1996)]{Brandner96} Brandner, W., Alcala,
J.~M., Kunkel, M., Moneti, A., \& Zinnecker, H.\ 1996, \aap, 307, 121

\bibitem[Chandler \& Carlstrom(1996)]{cc96} Chandler,
C.~J.~\& Carlstrom, J.~E.\ 1996, \apj, 466, 338

\bibitem[Cohen et al.(1981)]{Cohen}
Cohen, J.~G., Persson, S.~E., Elias, J.~H., \& Frogel, J.~A.\ 1981, \apj,
249, 481

\bibitem[Comer\'{o}n, Rieke, \& Rieke(1996)]{CRR96} Comeron, F.,
Rieke, G.~H., \& Rieke, M.~J.\ 1996, \apj, 473, 294

\bibitem[D'Antona \& Mazzitelli(1994)]{DM94} D'Antona, F.~\&
Mazzitelli, I.\ 1994, \apjs, 90, 467

\bibitem[D'Antona \& Mazzitelli(1997)]{DM98} D'Antona, F.~\&
Mazzitelli, I.\ 1997, Memorie della Societa Astronomica Italiana, 68, 807

\bibitem[Duquennoy \& Mayor(1991)]{DM91} Duquennoy, A.~\&
Mayor, M.\ 1991, \aap, 248, 485

\bibitem[Ghez, Neugebauer, \& Matthews(1993)]{GNM93} Ghez,
A.~M., Neugebauer, G., \& Matthews, K.\ 1993, \aj, 106, 2005

\bibitem[Ghez et al.(1997)]{Ghez97}
Ghez, A.~M., McCarthy, D.~W., Patience, J.~L., \& Beck, T.~L.\ 1997, \apj,
481, 378

\bibitem[Haisch et al.(2001)]{hai01} Haisch, K.~E., Lada,
E.~A., Pi{\~ n}a, R.~K., Telesco, C.~M., \& Lada, C.~J.\ 2001, \aj, 121,
1512

\bibitem[Haisch, Lada, \& Lada(2000)]{hai00} Haisch, K.~E.,
Lada, E.~A., \& Lada, C.~J.\ 2000, \aj, 120, 1396

\bibitem[Hester et al.(1996)]{Hester} Hester, J.~J.~et al.\
1996, \aj, 111, 2349

\bibitem[Hillenbrand \& Hartmann(1998)]{HH98} Hillenbrand,
L.~A.~\& Hartmann, L.~W.\ 1998, \apj, 492, 540

\bibitem[K{\" o}hler(2001)]{Kohler01} K{\" o}hler, R.\ 2001,
\aj, 122, 3325

\bibitem[K{\" o}hler \& Brandner(2001)]{KB01} K{\" o}hler,
R.~\& Brandner, W.\ 2001, IAU Symposium, 200, 147

\bibitem[Kroupa(2001)]{K01} Kroupa, P.\ 2001, \mnras, 322,
231

\bibitem[Kroupa, Aarseth, \& Hurley(2001)]{kah01} Kroupa, P.,
Aarseth, S., \& Hurley, J.\ 2001, \mnras, 321, 699

\bibitem[Kroupa, Petr, \& McCaughrean(1999)]{kpm99} Kroupa,
P., Petr, M.~G., \& McCaughrean, M.~J.\ 1999, New Astronomy, 4, 495

\bibitem[Kroupa, Tout, \& Gilmore(1993)]{KTG93} Kroupa, P.,
Tout, C.~A., \& Gilmore, G.\ 1993, \mnras, 262, 545

\bibitem[Lada et al.(1991)]{Lada91} Lada,
E.~A., Evans, N.~J., Depoy, D.~L., \& Gatley, I.\ 1991, \apj, 371, 171

\bibitem[Lada, Strom, \& Myers(1993)]{Lada93} Lada, E.~A.,
Strom, K.~M., \& Myers, P.~C.\ 1993, Protostars and Planets III, 245

\bibitem[Leinert et al.(1993)]{Leinert} Leinert, C., Zinnecker,
H., Weitzel, N., Christou, J., Ridgway, S.~T., Jameson, R., Haas, M., \&
Lenzen, R.\ 1993, \aap, 278, 129

\bibitem[Luhman et al.(2000)]{Luhman} Luhman, K.~L., Rieke,
G.~H., Young, E.~T., Cotera, A.~S., Chen, H., Rieke, M.~J., Schneider, G.,
\& Thompson, R.~I.\ 2000, \apj, 540, 1016

\bibitem[Mathieu et al.(2000)]{Mathieu}
Mathieu, R.~D., Ghez, A.~M., Jensen, E.~L.~N., \& Simon, M.\ 2000,
Protostars and Planets IV, 703

\bibitem[McLeod(1997)]{McLeod} McLeod, B.~A.\ 1997, The 1997
HST Calibration Workshop with a new generation of instruments /edited by
Stefano Casertano, Robert Jedrzejewski, Charles D.~Keyes, and Mark
Stevens.~Baltimore, MD : Space Telescope Science Institute (1997) QB
500.268 C35 1997, p.~281., 281

\bibitem[Mezger et al.(1988)]{mez88} Mezger, P.~G., Chini,
R., Kreysa, E., Wink, J.~E., \& Salter, C.~J.\ 1988, \aap, 191, 44

\bibitem[Mezger et al.(1992)]{Mezger} Mezger, P.~G., Sievers,
A.~W., Haslam, C.~G.~T., Kreysa, E., Lemke, R., Mauersberger, R., \&
Wilson, T.~L.\ 1992, \aap, 256, 631

\bibitem[Meyer(1996)]{Meyer_phd} Meyer, M.~R. \ 1996, Ph.D. Thesis,
University of Massachusetts

\bibitem[Meyer, Calvet, \& Hillenbrand(1997)]{Meyer97} Meyer,
M.~R., Calvet, N., \& Hillenbrand, L.~A.\ 1997, \aj, 114, 288

\bibitem[Meyer et al.(2000)]{Meyer_ppiv} Meyer, M.~R., Adams,
F.~C., Hillenbrand, L.~A., Carpenter, J.~M., \& Larson, R.~B.\ 2000,
Protostars and Planets IV, 121

\bibitem[Miller \& Scalo(1979)]{MS79} Miller, G.~E.~\&
Scalo, J.~M.\ 1979, \apjs, 41, 513

\bibitem[Mink(2002)]{mink} Mink, D.~J.\ 2002, ASP
Conf.~Ser.~281: Astronomical Data Analysis Software and Systems XI, 11, 169

\bibitem[Moore \& Yamashita(1995)]{moo95} Moore, T.~J.~T.~\&
Yamashita, T.\ 1995, \apj, 440, 722

\bibitem[Moore \& Chandler(1989)]{moo89} Moore, T.~J.~T.~\&
Chandler, C.~J.\ 1989, \mnras, 241, 19P

\bibitem[Najita, Tiede, \& Carr(2000)]{NTC00} Najita, J.~R.,
Tiede, G.~P., \& Carr, J.~S.\ 2000, \apj, 541, 977

\bibitem[Padgett, Strom, \& Ghez(1997)]{PSG97} Padgett,
D.~L., Strom, S.~E., \& Ghez, A.\ 1997, \apj, 477, 705

\bibitem[Petr et al.(1998)]{Petr} Petr, M.~G., Coude Du
Foresto, V., Beckwith, S.~V.~W., Richichi, A., \& McCaughrean, M.~J.\ 1998,
\apj, 500, 825

\bibitem[Reipurth \& Zinnecker(1993)]{RZ93} Reipurth, B.~\&
Zinnecker, H.\ 1993, \aap, 278, 81

\bibitem[Scalo(1986)]{Scalo86} Scalo, J.~M.\ 1986, Fundamentals
of Cosmic Physics, 11, 1

\bibitem[Simon, Close, \& Beck(1999)]{SCB99} Simon, M.,
Close, L.~M., \& Beck, T.~L.\ 1999, \aj, 117, 1375

\bibitem[Visser et al.(1998)]{viss}
Visser, A.~E., Richer, J.~S., Chandler, C.~J., \& Padman, R.\ 1998, \mnras,
301, 585

\end{thebibliography}
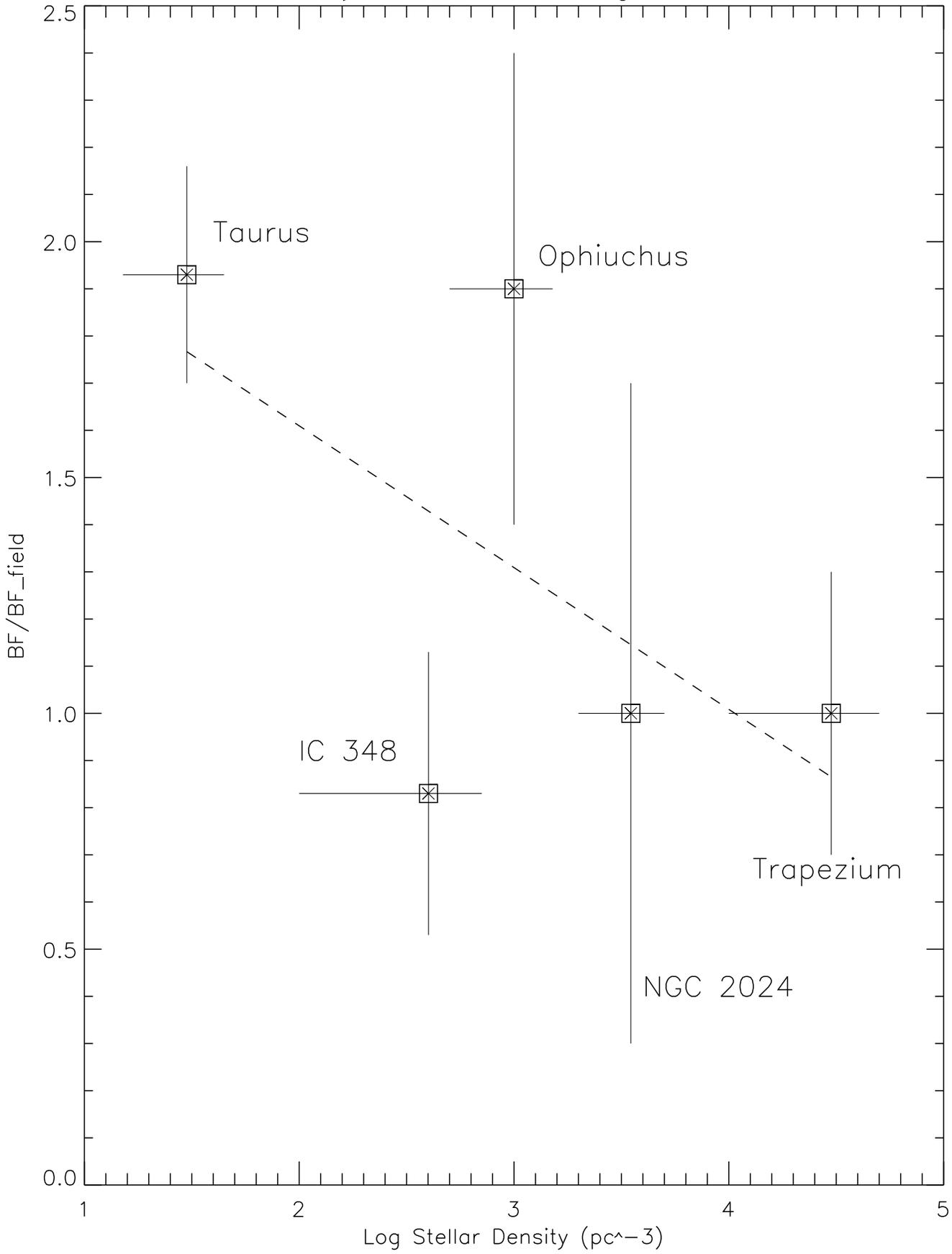
\end{document}